\documentstyle[12pt,psfig]{article}

\newcommand{\beq}{ \begin{eqnarray} }
\newcommand{\eeq}{ \end{eqnarray} }
\newcommand{\beqstar}{ \begin{eqnarray*} }
\newcommand{\eeqstar}{ \end{eqnarray*} }
\newcommand{\gsim}{ \mathop{}_{\textstyle \sim}^{\textstyle >} }
\newcommand{\lsim}{ \mathop{}_{\textstyle \sim}^{\textstyle <} }

\begin{document}
\baselineskip 0.7cm

\begin{titlepage}

\begin{center}

\hfill CERN-TH/2001-057\\
\hfill KEK-TH-749\\
\hfill February 2001\\
\hfill hep-ph/yymmdd

  {\large Neutrino masses, muon $g-2$, and 
  lepton-flavour violation\\
  in the supersymmetric see-saw model} 
\vskip 0.5in 
{\large Junji~Hisano$^{(a,b)}$ and Kazuhiro~Tobe$^{(b)}$}
\vskip 0.4cm 
{\it 
(a) Theory Group, KEK, Oho 1-1, Tsukuba, Ibaraki 305-0801, Japan
}
\\
{\it 
(b) Theory Division, CERN, 1211 Geneva 23, Switzerland
}
\vskip 0.5in

\abstract {
  In the light of the recent muon $(g_\mu-2)$ result by the E821
  experiment at the Brookhaven National Laboratory, we study the event
  rates of the charged lepton-flavour-violating (LFV) processes in the
  supersymmetric standard model (SUSY SM) with the heavy right-handed
  neutrinos (SUSY see-saw model). Since the left-handed sleptons get
  the LFV masses via the neutrino Yukawa interaction in this model,
  the event rate of $\mu\rightarrow e \gamma$ and the SUSY-SM
  correction to $(g_\mu-2)/2$ ($\delta a_\mu^{\rm SUSY}$) are strongly
  correlated.  When the left-handed sleptons have a LFV mass between
  the first and second generations ($(m^2_{\tilde L})_{12}$) in the
  mass matrix, it should be suppressed by $\sim 10^{-3}$
  $({10^{-9}}/{\delta a_\mu^{\rm SUSY}})$ compared with the diagonal
  components ($m_{\rm SUSY}^2$), from the current experimental bound
  on $\mu\rightarrow e \gamma$. The recent $(g_\mu-2)$ result
  indicates ${\delta a_\mu^{\rm SUSY}}\sim 10^{-9}$. The future
  charged LFV experiments could cover $(m^2_{\tilde L})_{12}/m_{\rm
  SUSY}^2\gsim 10^{-(5-6)}$.  These experiments will give a
  significant impact on the flavour models and the SUSY-breaking
  models.  In the SUSY see-saw model $(m^2_{\tilde L})_{12}$ is
  proportional to square of the tau-neutrino Yukawa-coupling constant.
  In the typical models where the neutrino-oscillation results are
  explained and the top-quark and tau-neutrino Yukawa couplings are
  unified at the GUT scale, a large LFV mass of $(m^2_{\tilde
  L})_{12}/m_{\rm SUSY}^2\gsim 10^{-4}$ is generated, and the large
  LFV event rates are predicted.  We impose a so-called no-scale
  condition for the SUSY-breaking parameters at the GUT scale, which
  suppresses the FCNC processes, and derive the conservative lower bound
  on $\mu\rightarrow e\gamma$.  The predicted Br($\mu\rightarrow e
  \gamma$) could be covered at the future LFV experiments.  }
\end{center}
\end{titlepage}
\setcounter{footnote}{0}

Various neutrino oscillation experiments suggest the existence of
lepton-flavour violation (LFV). The atmospheric neutrino result by the
superKamiokande experiment is a convincing evidence of the neutrino
mixing between $\nu_\mu$ and $\nu_\tau$ \cite{atm}.  Solar neutrino
experiments have also strongly indicated the mixing between $\nu_e$
and $\nu_\mu$ or $\nu_\tau$ \cite{solar}.  These results are explained
by the tiny but non-zero neutrino masses, and the natural model for
the neutrino masses is the see-saw mechanism \cite{seesaw}.

It has been discussed that the supersymmetric (SUSY) extension of the
see-saw mechanism (SUSY see-saw model) could lead to the charged-LFV
processes, such as $\mu\rightarrow e \gamma$
\cite{Borzumati:1986qx,Hisano:1996cp,Sato:2001zh,
Hisano:1999fj,others}.  If the SUSY-breaking terms in the
supersymmetric standard model (SUSY SM) are generated by the
interaction at the gravitational scale or the GUT scale, the Yukawa
interaction of the right-handed neutrinos generates the LFV masses for
the left-handed sleptons radiatively, and the LFV masses are a direct
source of the charged-LFV processes in the SUSY SM. Especially, in the
SUSY models with Yukawa-coupling unification between the top quark and
tau neutrino, the event rates of the charged-LFV processes are
significantly enhanced, since the LFV masses is proportional to square
of the tau-neutrino Yukawa-coupling constant.  In many proposed
models, this assumption is adopted so that the fermion mass structure,
including the neutrino sector, is explained in a unified picture
between quarks and leptons, of which a well-known example is the
SO(10) GUT. If the left-handed slepton, chargino and neutralino are
light, the predicted event rates could be within reach of the
near-future experiments.

Recently, the ongoing E821 experiment at the Brookhaven National
Laboratory updated the result for the muon anomalous magnetic moment
$(g_\mu-2)$ \cite{Brown:2001mg}, and it is found that $(g_\mu-2)$ is
about $2.6\sigma$ away from the SM prediction as
\begin{eqnarray}
a_\mu({\rm exp})-a_\mu({\rm SM})=43(16) \times 10^{-10}.
\label{g_2_exp}
\end{eqnarray}
This suggests that new physics exists around the TeV scale
\cite{Czarnecki:2001pv}, and this result will be further refined after
including the data for the 2000 run.  The SUSY SM predicts a sizable
deviation from the SM prediction for $(g_\mu-2)$ \cite{g-2_before},
and it is a good candidate to accommodate the $(g_\mu-2)$ result
\cite{g-2_recent}. This comes from the fact that the dominant
correction to $(g_\mu-2)$ in the SUSY SM is from a one-loop diagram of
chargino and muon sneutrino and it is proportional to $\tan\beta$.
This new observation indicates that the chargino, the left-handed
smuon and the sneutrino are relatively light or that $\tan\beta$ is
much larger than 1.

In this situation the LFV processes of muon will be extremely
interesting, since the event rates for the LFV processes are enhanced
by the light chargino and slepton or large $\tan\beta$.\footnote{
In Ref.~\cite{g-2lfv1} the relation between $(g_\mu-2)$ and the
charged LFV processes is presented in non-supersymmetric models to
generate finite neutrino masses. Also, in Ref.~\cite{g-2lfv2} the
relation is discussed in the $Z'$ model with the flavor-dependent
interaction.}
 In fact, when the left-handed sleptons have LFV masses, the 1-loop
diagrams in $\mu\rightarrow e\gamma$ have the same structure as those
in the SUSY-SM correction to $(g_\mu-2)$. As a result, the event rate
for $\mu\rightarrow e\gamma$ is approximately proportional to the
square of the SUSY-SM correction to $(g_\mu-2)$.  The other LFV
processes of muon, $\mu\rightarrow 3e$ and $\mu-e$ conversion in
nuclei, are also enhanced since the photon penguin diagram tends to
dominate in those processes.

In this letter, in the light of the recent $(g_\mu-2)$ result, we
derive the conservative event rates of the charged-LFV processes in
the SUSY see-saw models with Yukawa-coupling unification. In order to
derive the conservative prediction of the LFV event rates, we consider
a so-called no-scale-type SUSY breaking at the GUT scale
\cite{Ellis:1984bm} (or recently called  gaugino mediation
\cite{Kaplan:2000ac}). This means that all SUSY-breaking scalar masses
and A-terms vanish at the GUT scale, and hence the dominant source of SUSY
breaking is non-zero gaugino masses. This no-scale-type SUSY breaking
is well-known as a suppression mechanism of the SUSY FCNC
problem. Since the $(g_\mu-2)$ result gives the upper bounds of the
slepton and chargino masses, this limit supplies the conservative
lower bound on the LFV event rates.  We show that the proposed
improvement of limits on the LFV rare muon decay processes and the
present indication of anomaly for $(g_\mu-2)$ will provide a
significant impact on the flavour models as well as the SUSY-breaking
mechanism.

Before starting to discuss the charged-LFV processes in the SUSY SM
with the right-handed neutrinos, we show the strong correlation
between the event rate of $\mu\rightarrow e \gamma$ and the
SUSY-SM correction to $(g_\mu-2)$ when the left-handed sleptons have
the LFV masses.  These are generated by one-loop diagrams mediated by
sleptons, neutralinos and charginos. The effective operators for
$\mu\rightarrow e \gamma$ and $(g_\mu-2)$ are
\begin{eqnarray}
{\cal L}_{eff} = 
e \frac{m_{l_j}}{2}\bar{l}_i \sigma_{\mu\nu}  
F^{\mu\nu} (L_{ij} P_L+R_{ij} P_R) l_j
\label{eff_op}
\end{eqnarray}
where $m_{l_i}$ is a mass of the charged lepton $l_i$, $e$ and
$F^{\mu\nu}$ are the QED coupling constant and field strength, and
$P_{R/L}=(1\pm \gamma_5)/2$. The coefficients $L_{ij}$ and $R_{ij}$
are functions of the SUSY-particle masses and mixings. From these
operators (\ref{eff_op}), the branching ratio for $\mu \rightarrow e
\gamma$ is given by
\begin{eqnarray}
{\rm Br}(\mu \rightarrow e \gamma) 
= 
\frac{48\pi^3 \alpha}{G_F^2} (|L_{12}|^2+|R_{12}|^2),
\end{eqnarray}
and the SUSY-SM correction to $a_\mu(\equiv (g_\mu-2)/2)$ is 
\begin{eqnarray}
\delta a_\mu^{\rm SUSY}=m_\mu^2 (L_{22}+R_{22}).
\end{eqnarray}
The chargino--sneutrino diagram tends to dominate over the other
diagrams in the SUSY-SM contribution to $(g_\mu-2)$.  Assuming all
SUSY particle masses are the same, the SUSY-SM correction is
\begin{eqnarray}
\delta a_\mu^{\rm SUSY} \simeq
\frac{5\alpha_2^2+\alpha_Y^2}{48\pi}\frac{m_\mu^2}{m^2_{\rm SUSY}}\tan\beta
\label{amuap}
\end{eqnarray}
for $\tan\beta \gsim 1$. Here, $L_{22}\simeq R_{22}$.  When the
left-handed sleptons have the LFV masses between the first and second
generations ($(m^2_{\tilde L})_{12}$), the chargino--sneutrino diagram
also dominates in $\mu \rightarrow e \gamma$. In this case,
$|R_{12}|\gg|L_{12}|$.  Taking all SUSY-particle masses common again,
we get
\begin{eqnarray}
{\rm Br}(\mu \rightarrow e \gamma) 
&\simeq& 
\frac{\pi}{75} \alpha (\alpha_2+\frac{5}{4}\alpha_Y)^2 
(G_F^2 m_{\rm SUSY}^4)^{-1} \tan^2\beta 
\left(\frac{(m^2_{\tilde L})_{12}}{m_{\rm SUSY}^2}\right)^2
\nonumber\\
&=&
3 \times 10^{-5} 
\left(\frac{\delta  a_\mu^{\rm SUSY}}{10^{-9}}\right)^2
\left(\frac{(m^2_{\tilde L})_{12}}{m_{\rm SUSY}^2}\right)^2.
\label{bmuap}
\end{eqnarray}
The value $\delta a_\mu^{\rm SUSY}>10^{-9}$ is favored by the
recent result from the E821 experiment at the $2 \sigma$ level.  From the
current experimental bound of $\mu\rightarrow e\gamma$ (Br($\mu
\rightarrow e \gamma$)$<1.2\times 10^{-11}$ \cite{Brooks:1999pu}),
this imposes a stringent constraint on the LFV mass for the
left-handed sleptons as\footnote{
When only the right-handed sleptons have the LFV masses, the
correlation between ${\rm Br}(\mu \rightarrow e \gamma)$ and
$(g_\mu-2)$ is relatively weak.  This is because the dominant
contribution to $\mu\rightarrow e \gamma$ comes from bino--slepton
diagrams, while bino--slepton diagrams are subdominant in $(g_\mu-2)$.
Moreover, there is an accidental cancellation among the diagrams in
$\mu \rightarrow e \gamma$ \cite{Hisano:1997qq}. If both the
left-handed and right-handed sleptons have the LFV masses, the
correlation depends on the LFV masses and the other SUSY-breaking
parameters. In this case, the event rate of $\mu \rightarrow e \gamma$
is determined by the competition between a neutralino--slepton diagram
proportional to $m_\tau$ and the chargino-sneutrino diagram
\cite{Okada:2000zk}.
}
\begin{eqnarray}
\frac{(m^2_{\tilde L})_{12}}{m_{\rm SUSY}^2}
\lsim 6 \times10^{-4} 
\left(\frac{\delta a_\mu^{\rm SUSY}}{10^{-9}}\right)^{-1}
\left(
\frac{{\rm Br}(\mu \rightarrow e \gamma)}{1.2\times 10^{-11}}\right)^{\frac12}.
\label{minus5}
\end{eqnarray}

This constraint (\ref{minus5}) means that the off-diagonal component
in the slepton mass matrix should be much smaller than the diagonal
ones, so that $\mu \rightarrow e \gamma$ is suppressed. On the other
hand, the LFV mass terms for sleptons are not forbidden by any
symmetries in the presence of the neutrino masses. The decoupling
solution (so-called effective SUSY model) has been proposed as a
solution for the SUSY FCNC problem \cite{effsusy}. In this model the
slepton and squark masses for the first and second generations are so
heavy in order to suppress the FCNC processes, while the off-diagonal
components in the squark and slepton mass matrixes are comparable with
the diagonal ones. Now the effective SUSY model is disfavoured, since
the constraint (\ref{minus5}) is independent of the SUSY-breaking
scale $m_{\rm SUSY}$ and $\tan\beta$. The above result supports the
universal scalar mass hypothesis for the SUSY breaking parameters,
such as in the minimal supergravity scenario \cite{Nilles:1984ge} or
the gauge-mediated SUSY-breaking scenario \cite{gmsb}. In this
hypothesis, the charged-LFV processes are suppressed by the super-GIM
mechanism.  However, if the scale for the generation of the
SUSY-breaking terms is higher than the right-handed neutrino mass
scale in the SUSY see-saw model, the renormalization effect could
induce the sizable off-diagonal components. The future charged-LFV
experiments could cover $(m^2_{\tilde L})_{12}/m_{\rm SUSY}^2\gsim
10^{-(5-6)}$, as will be shown. Thus, these experiments will give a
significant impact on the flavour models and the SUSY-breaking models.

Let us start to discuss the neutrino masses and mixing matrix in the
SUSY SM with right-handed neutrinos and present the
charged-LFV event rates, using the result for the neutrino oscillation
experiment and $(g_\mu-2)$. In the SUSY SM with the right-handed
neutrinos ($\bar{N}_i$), the superpotential in the lepton sector is
given by
\begin{eqnarray}
W &=& \bar{E}_i f_e^i L_i H_d + \bar{N}_i f_\nu^{ij} L_j H_u
+\frac{1}{2} \bar{N}_i M_{Ri} \bar{N}_i,
\end{eqnarray}
where $\bar{E}_i$ and $L_i$ are right-handed charged leptons and
lepton doublets, respectively. Without loss of generality, we can take
a basis in which the charged-lepton Yukawa coupling matrix $f_e$ and
the right-handed neutrino mass matrix $M_{R}$ are diagonalized. In
this basis the neutrino Yukawa coupling matrix $f_\nu$ has flavour
mixings generically, and this is the origin of LFV.  After integrating
out the heavy right-handed neutrinos, the neutrino mass matrix is given
as
\begin{eqnarray}
m_\nu =m_{\nu D}^T M_R^{-1} m_{\nu D}.
\end{eqnarray}
Here, the Dirac neutrino masses $m_{\nu D}$ are given by $m_{\nu
  D}=f_\nu \langle H_u \rangle$.  The neutrino mixing matrix $V_{MNS}$
is defined by
\begin{eqnarray}
V_{MNS}^T m_\nu V_{MNS} = {\rm diag}(m_{\nu 1},~m_{\nu 2},~m_{\nu 3}).
\end{eqnarray}
The tiny neutrino masses and non-zero neutrino mixings provide
neutrino oscillation solutions to atmospheric and solar neutrino
problems.

In order to explain the atmospheric neutrino result, we need an almost 
maximal mixing between $\nu_\mu$ and $\nu_\tau$ ($(V_{MNS})_{\mu3}
\simeq 1/\sqrt{2}$). Thus, the structure of the neutrino mixing matrix
looks very different from that of the Kobayashi--Maskawa mixing matrix in
the quark sector. So far many models have been proposed in order to
naturally accommodate both small mixings in the quark sector and large
mixing in the neutrino sector within the context of unified theories.  An
interesting proposal is that a large mixing for neutrinos can be
easily understood by a lopsided structure of the Dirac mass matrices
(that is, $f_\nu^{22}\sim f_\nu^{23}\ll f_\nu^{33}\sim f_\nu^{32}$)
without introducing a non-trivial structure of the right-handed
neutrino mass matrix \cite{neutrino_models1}. In this case, the lepton
mixing matrix $V$, which diagonalizes the neutrino Yukawa matrix as
\begin{eqnarray}
U f_\nu V ={\rm diag} (f_{\nu 1},~f_{\nu 2},~f_{\nu 3}),
\end{eqnarray}
also has a large mixing between the second and third generations,
i.e. $V_{23} \sim 1/\sqrt{2}$. The component $V_{13}$ is
model-dependent; however, it tends to be larger than 0.01.  We list
typical predicted values for $V_{23}$ and $V_{13}$ in some proposed
models in Table \ref{list}.  The order-one $V_{23}$ and non-zero
$V_{13}$ elements are very important for the charged-LFV processes in
SUSY models, as stressed in Refs.~\cite{Sato:2001zh}. The component
$V_{12}$ depends on the solution of the solar neutrino problem. It may
also give a sizable contribution to the LFV processes if $f_{\nu 2}$
and $V_{12}$ are relatively large \cite{Hisano:1999fj}.  However, we
take a limit of vanishing $f_{\nu 2}$ in order to derive the
conservative predictions in this paper.

\begin{table}
\begin{center}
\begin{tabular}{|c|c|c|}\hline
Models & $V_{23}$ & $V_{13}$  \\
\hline 
Albright {\it et al.} \cite{neutrino_models2} & 0.9 &  0.06 \\
Altarelli {\it et al.} \cite{neutrino_models2} & 0.5 & 0.09 \\
Bando {\it et al.} \cite{neutrino_models2} 
& $\sim 0.7$ & $\sim 0.1$  \\
Hagiwara {\it et al.} \cite{neutrino_models2} & 0.7 & 0.06  \\
Nomura {\it et al.} \cite{neutrino_models2} & 0.7 & $\sim 0.1$  \\
Sato {\it et al.} and Buchm\"uller {\it et al.} 
\cite{neutrino_models1, JY}& 0.7 & $\sim 0.05$ \\
\hline
\end{tabular}
\end{center}
\caption{Typical predicted values for $V_{23}$ and $V_{13}$
in various models \cite{neutrino_models1, JY, neutrino_models2}.}
\label{list}
\end{table}

Using the above results for the neutrino mixing matrix, we discuss the
charged-LFV processes in the SUSY SM with the right-handed neutrinos.
In order to describe a real world, soft SUSY-breaking terms have to be
included in the SUSY SM. As mentioned above, the current situation
favours the universal scalar mass hypothesis for the SUSY-breaking
parameters. If the scale for the generation of the SUSY-breaking terms
is higher, the existence of the large neutrino Yukawa coupling can
induce significant LFV masses through the renormalization effect, as
pointed out in Refs.~\cite{Borzumati:1986qx,Hisano:1996cp}. In the
following, the scale for the generation of the SUSY-breaking terms is
assumed to be the GUT scale or the reduced Planck scale, as in the
minimal supergravity scenario. Since the left-handed leptons couple to
the right-handed neutrinos, LFV is induced in left-handed slepton
masses.

The renormalization-group equation (RGE) for the left-handed slepton
masses is given by
\begin{eqnarray}
&&\mu \frac{d}{d\mu} (m^2_{\tilde{L}})_{ij} =
\left(\mu \frac{d}{d\mu} (m^2_{\tilde{L}})_{ij} \right)_{\rm SUSY\mbox{-}SM}
\nonumber
\\
&+&\frac{1}{16 \pi^2} \left\{ m^2_{\tilde{L}} f^\dagger_\nu f_\nu
+f^\dagger_\nu f_\nu m^2_{\tilde L}+
2\left(f^\dagger_\nu m^2_{\tilde \nu} f_\nu +m^2_{H_u} f^\dagger_\nu f_\nu
+A^\dagger_\nu A_\nu \right) \right\}_{ij},
\label{RGE}
\end{eqnarray}
where $m_{\tilde L}$, $m_{\tilde \nu}$, $m_{H_u}$, and $A_\nu$
denote the SUSY-breaking masses for the doublet slepton $({\tilde L})$,
the right-handed sneutrino $({\tilde \nu})$, the Higgs $(H_u)$, and
the trilinear A-term for sneutrinos, respectively.
Here, $(\mu \frac{d}{d\mu} (m^2_{\tilde{L}})_{ij} )_{\rm SUSY\mbox{-}SM}$ 
represents the RGE in the SUSY SM:
\begin{eqnarray}
\left(\mu \frac{d}{d\mu} (m^2_{\tilde{L}})_{ij} \right)_{\rm SUSY\mbox{-}SM}
&=& \frac{1}{16 \pi^2} \left\{
-\left(\frac{6}{5} g_1^2 M_1^2 + 6 g^2_2 M^2_2 \right) \delta_{ij}
\right. \nonumber \\
&&+ (m^2_{\tilde L} f^\dagger_e f_e +f^\dagger_e f_e m^2_{\tilde L})_{ij}
\nonumber \\
&& \left. +2 (f_e^\dagger m^2_{\tilde e} f_e + m^2_{H_d} f_e^\dagger f_e 
+A_e^\dagger A_e )_{ij} \right\},
\label{RGE_MSSM}
\end{eqnarray}
where $M_1$, $M_2$, $m_{\tilde e}$, $m_{H_d}$ and $A_e$ denote the bino
mass, wino mass, SUSY-breaking masses for the charged slepton
$({\tilde e})$, the Higgs $(H_d)$, and the trilinear A-term for charged
sleptons, respectively. 

If we assume the universal scalar mass $(m_0)$ for all scalar bosons
and the universal A-term $(A_f=a_0 m_0 f_f)$ at the unification scale
($M_G=2\times 10^{16}$ GeV), the only source of LFV is the neutrino
Yukawa coupling $f_\nu$.  An iteration gives an approximate solution
to Eq.~(\ref{RGE}) for the diagonal and off-diagonal components of the
left-handed slepton mass matrix:
\begin{eqnarray}
(m^2_{\tilde{L}})_{ii} &\simeq& m_0^2 +\frac{1}{16\pi^2}
\left\{\frac{36g_G^2 M_0^2}{5} 
-m^2_0(6+a_0^2) \left|V_{i3} f_{\nu 3} \right|^2 \right\}
\log \frac{M_G}{M_R}, \nonumber
\\
(m^2_{\tilde{L}})_{ij} &\simeq& -\frac{(6+a_0^2)m^2_0}
{16 \pi^2} V_{i3} V_{j3}^* |f_{\nu 3}|^2 \log \frac{M_G}{M_R},
~~~{\rm for}~~i\neq j.
\label{LFV_slepton}
\end{eqnarray}
Here we also assumed the gaugino mass unification ($M_0$) at $M_G$,
and the hierarchy $f_{\nu 3}\gg f_{\nu 2} \gg f_{\nu 1}$. In this
approximation, we neglected all charged-lepton Yukawa couplings $f_e$.
As can be seen in Eqs.~(\ref{LFV_slepton}), the large neutrino Yukawa
couplings $f_{\nu3}$ and mixing $V$ induce large LFV masses as well as
non-degeneracy in diagonal components of the slepton mass matrix. If
$f_{\nu3}= f_{top}$ at the GUT scale and $m_0 \sim M_0$, $(m^2_{\tilde
  L})_{12}/m^2_{\rm SUSY}\sim 10^{-1}\times V_{23}V_{13}$. From
Eq.~(\ref{minus5}), it is found that the models with $V_{23}\simeq
1/\sqrt{2}$ and $V_{13}\gsim 0.01$ are marginal.  When $m_0$ is much
smaller than $M_0$, the LFV masses $(m^2_{\tilde L})_{ij}~(i\neq j)$
are suppressed with respect to the diagonal elements $(m^2_{\tilde
  L})_{ii}$, and all the diagonal elements tend to be degenerate.  In
the limit of $m_0=0$ (the no-scale limit), the LFV masses are
generated only at higher order.  In these cases the charged-LFV
processes are suppressed while the correction to the $(g_\mu-2)$ is
not suppressed.  However, the large neutrino Yukawa coupling still
induces the significant effect, even if we take $m_0=0$ at $M_G$, as
will be shown.

\begin{figure}[p]
\centerline{\psfig{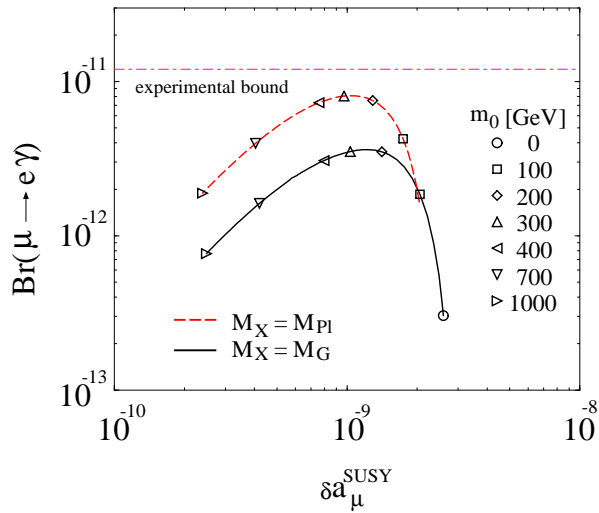}}
\caption{Br($\mu \rightarrow e \gamma$) and $\delta a_\mu^{\rm SUSY}$
  as a function of universal mass $m_0$. Here we take $V_{13}=0.01$,
  $V_{23}=1/\sqrt{2}$, $\tan\beta=10$ and $M_2=250$ GeV. We assume
  unification between the top-quark and tau-neutrino Yukawa
  couplings ($m_t=175$ GeV and $m_{\nu_\tau}=0.055$ eV). The solid and
  dashed lines are for cases where the scale for the generation of the
  SUSY-breaking terms in the SUSY SM ($M_X$) are the GUT scale and the
  reduced Planck scale, respectively.  }
\label{slep_dep}
\end{figure}

As listed in Table~\ref{list}, many models for realistic fermion
masses imply $V_{23}\simeq 1/\sqrt{2}$ and $V_{13}>0.01$. Therefore,
we first fix $V_{23}=1/\sqrt{2}$ and $V_{13}=0.01$, and show the
predicted Br($\mu \rightarrow e \gamma$) and $\delta a_\mu^{\rm SUSY}$
in Fig.~{\ref{slep_dep}}. Here, unification between the top-quark and
tau-neutrino Yukawa coupling constants ($f_{\nu3}= f_{top}$) at the
GUT scale is assumed. We take $\tan \beta=10$, $M_2=250$ GeV,
$m_t=175$ GeV and $m_{\nu_\tau}=0.055$ eV.  In this case the
tau-neutrino Yukawa coupling $f_{\nu3}$ at the GUT scale is 0.58 and
$M_R$ is $2\times 10^{14}$ GeV and. In our analysis, we numerically
solve the RGEs for couplings and SUSY-breaking parameters, and
calculate the branching ratios of $\mu \rightarrow e \gamma$ and
$(g_\mu-2)$ by using the complete formula given in
Ref.~\cite{Hisano:1996cp}.

The solid line is for a case where the scale for generation of the
SUSY-breaking terms in the SUSY SM ($M_X$) is the GUT scale. From this
figure, a smaller $m_0$ region is favored to explain the recent
$(g_\mu-2)$ result given in Eq.~(\ref{g_2_exp}). When $m_0$ is larger,
both $\delta a_\mu^{\rm SUSY}$ and Br($\mu\rightarrow e\gamma$) are
suppressed as in Eqs.~(\ref{amuap}) and (\ref{bmuap}). On the other hand, in
the region with $m_0 \ll M_2$, only Br($\mu \rightarrow e \gamma$) is
suppressed, as explained above. However, the large neutrino Yukawa
coupling still induces a sizable effect on Br($\mu \rightarrow e
\gamma$), even in the no-scale limit ($m_0=0$).

So far we assumed that the SUSY-breaking terms are generated at the GUT
scale. In fact, this leads to the conservative bound on the
$\mu\rightarrow e \gamma$ event rate. If $M_X$ is larger than the GUT
scale, the running effect above this scale also induces LFV masses
in general. Therefore, the branching ratios for the charged-LFV
processes will become much larger, and our result for $M_X=M_G$ is
conservative.  In Fig.~\ref{slep_dep}, the dashed line is for the case
where $M_X$ is the reduced Planck scale ($M_{Pl}=2.4\times 10^{18}$
GeV). Here, we use the RGE of the SUSY SM with the right-handed
neutrino even above the GUT scale, while we assume the GUT relation of
the gaugino masses for simplicity.  Since the ratio of $M_X$ to $M_R$
is larger, the event rate becomes larger.

\begin{figure}[p]
\centerline{\psfig{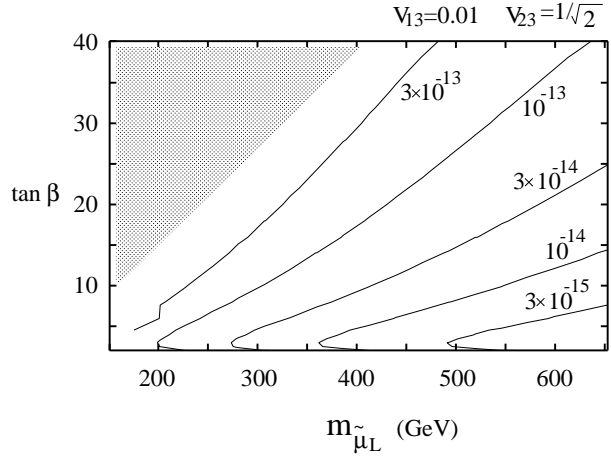}}
\centerline{\psfig{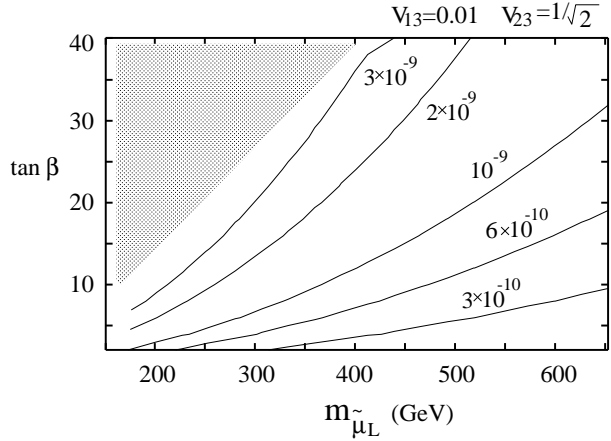}}
\caption{(a) Br($\mu \rightarrow e \gamma$) and (b) $\delta a_\mu^{\rm
    SUSY}$ as functions of the left-handed smuon mass and $\tan\beta$.
  Here, we impose the no-scale condition ($m_0=0$) at the GUT scale in
  order to derive the conservative lower bound on Br($\mu \rightarrow
  e \gamma$). We take $V_{13}=0.01$ and $V_{23}=1/\sqrt{2}$, and
  assume unification between the top-quark and tau-neutrino Yukawa
  couplings as in Fig.~\ref{slep_dep}.  }
\label{tanbeta_dep}
\end{figure}

In order to obtain a conservative lower bound on Br($\mu \rightarrow
e \gamma$) below, we take the no-scale-type initial condition $m_0=0$ at
$M_G$, and show Br($\mu \rightarrow e \gamma$) and $\delta a_\mu^{\rm
SUSY}$ as a function of the left-handed smuon mass and $\tan \beta$ in
Fig.~\ref{tanbeta_dep}.\footnote{
  If we seriously take the no-scale-type initial condition at the GUT
  scale, the slepton may be the LSP, and hence we may need a lighter
  SUSY particle, such as the axino. Otherwise, if we take into account
  the running effect above the GUT scale, which is highly dependent on
  GUT models, this problem would be solved \cite{Komine:2001tj}. The
  detailed studies for this model, including $b\rightarrow s\gamma$,
  the light Higgs boson mass, and the constraint from the negative
  SUSY searches, are done in Refs.~\cite{komine}, and the analysis
  shows that the region with $M_2\lsim180$GeV or $\tan\beta
  \lsim(2-5)$ is excluded if the SUSY breaking terms are generated at
  the GUT scale.  }
Here we take $V_{13}=0.01$ and $V_{23}=1/\sqrt{2}$.  As the SUSY
contribution to $(g_\mu-2)$ increases, the branching ratio of $\mu
\rightarrow e \gamma$ gets larger. In the parameter region with
$\delta a_\mu^{\rm SUSY}>10^{-9}$, which is favored by the recent
result at the $2 \sigma$ level, Br($\mu \rightarrow e \gamma$) is
larger than $10^{-14}$.  This interesting region will be covered by a
near-future $\mu \rightarrow e \gamma$ experiment at PSI \cite{PSI} if
their proposed goal of $10^{-14}$ is achieved.

\begin{figure}
\centerline{\psfig{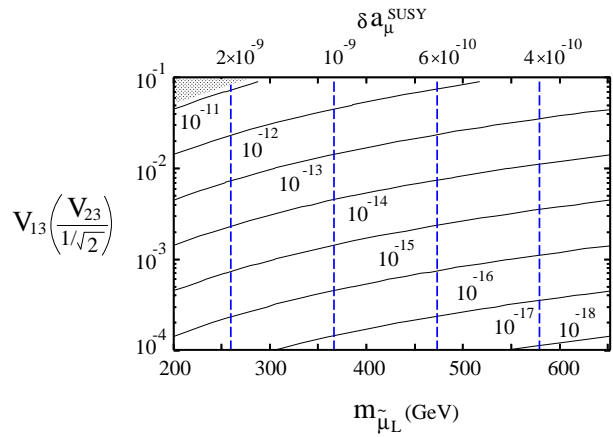}}
\caption{Br($\mu \rightarrow e \gamma$) as a function of $V_{13}
  V_{23}$ and the left-handed smuon mass or $\delta a_\mu^{\rm SUSY}$
  assuming $\tan\beta=10$.  We impose the no-scale condition ($m_0=0$)
  at the GUT scale.  Unification between the top-quark and
  tau-neutrino Yukawa couplings is assumed as in Fig.~\ref{slep_dep}.  In
  typical models, $V_{13}\gsim 10^{-2}$, as listed in Table~\ref{list}.
}
\label{v13_dep}
\end{figure}

In Fig.~\ref{v13_dep}, we show Br($\mu \rightarrow e \gamma$) as a
function of $V_{13} V_{23}$ and the left-handed smuon mass or $\delta
a_\mu^{\rm SUSY}$ assuming $\tan\beta=10$.  At present the
experimental limit Br($\mu \rightarrow e \gamma$)$<1.2\times 10^{-11}$
is not very significant.  However, the future expected reach at PSI
(Br($\mu \rightarrow e \gamma$)$\sim 10^{-14})$ will be quite
significant. Since, from Fig.~\ref{tanbeta_dep}, the region favored by
the present $(g_\mu-2)$ experiment is about $m_{\tilde{L}}< 360$ GeV
for $\delta a_\mu^{\rm SUSY} <10^{-9}$, so the future limit on lepton
mixing $V_{13}(\frac{V_{23}}{1/\sqrt{2}})$ will be about $4\times
10^{-3}$.  This limit will have an impact on the fermion mass models
as listed in Table~\ref{list}.  We should stress again that in
Figs.~\ref{tanbeta_dep} and \ref{v13_dep}, we took the no-scale-type
initial condition ($m_0=0$) at the GUT scale, which significantly
suppress the flavour violation in the SUSY-breaking parameters as can
be seen in Fig.~\ref{slep_dep}.  So if we consider non-zero universal
mass ($m_0\neq 0$) or if $M_X$ is larger than $M_G$, the branching
ratio of $\mu \rightarrow e \gamma$ becomes larger.

Finally, we comment on another interesting process, the $\mu$--$e$
conversion in nuclei. In the SUSY see-saw model, the photon penguin
diagram tends to dominate over the other contributions in the $\mu$--$e$
conversion process in a wide range of the parameter space.  Therefore,
we have the following relation between event rates Br($\mu \rightarrow e
\gamma$) and R($\mu \rightarrow e$ in nuclei):
\begin{eqnarray}
\frac{{\rm R}(\mu \rightarrow e~{\rm in~Ti~(Al)})}
{{\rm Br}(\mu \rightarrow e \gamma)} \simeq 5~(3)\times 10^{-3}.
\label{mec}
\end{eqnarray}
The proposed experiment MECO at BNL will reach $10^{-16}$ for the
conversion rate in Al \cite{MECO}. From relation (\ref{mec}), it is
found that the MECO experiment will also provide a significant probe
on LFV in the SUSY SM.  Furthermore, according to ongoing studies for
the $\nu$ factory at CERN \cite{CERN} and the PRISM project at
KEK/JAERI \cite{PRISM}, a sensitivity of $10^{-18}$ for R$(\mu
\rightarrow e~{\rm in~Ti})$ may be possible. A value of R$(\mu
\rightarrow e~{\rm in~Ti})\lsim10^{-18}$ corresponds to Br$(\mu
\rightarrow e \gamma)\lsim 10^{-16}$, therefore $V_{13}\lsim
10^{-(3-4)}$. If this is achieved and the anomaly of $(g_\mu-2)$ is also
well-established in the future, not only realistic fermion-mass models but
also the SUSY-breaking mechanism under the universal scalar mass
hypothesis, including no-scale-type (or gaugino-mediation) SUSY
breaking, will be significantly probed by the LFV and $(g_\mu-2)$
experiments.

%%%%%%%%%%%%%%%%%%%%%%%%%%%
\section*{Acknowledgement}
This work was also supported in part by the Grant-in-Aid for Scientific
Research from the Ministry of Education, Science, Sports and Culture
of Japan, on Priority Area 707 ``Supersymmetry and Unified Theory of
Elementary Particles" (J.H.).
%%%%%%%%%%%%%%%%%%%%%%%%%%%


\begin{thebibliography}{99}
%
\bibitem{atm}
Y.~Fukuda {\it et al.}  [Super-Kamiokande Collaboration],
Phys.\ Rev.\ Lett.\ {\bf 81}, 1562 (1998)
[hep-ex/9807003];
{\it ibid} \ {\bf 85}, 3999 (2000)
[hep-ex/0009001].

%
\bibitem{solar}
R.~J.~Davis, D.~S.~Harmer and K.~C.~Hoffman,
Phys.\ Rev.\ Lett.\ {\bf 20}, 1205 (1968);\\
Y.~Fukuda {\it et al.}  [Kamiokande Collaboration],
Phys.\ Rev.\ Lett.\ {\bf 77}, 1683 (1996);\\
K.~Lande {\it et al.},
Astrophys.\ J.\ {\bf 496}, 505 (1998);\\
D.~N.~Abdurashitov {\it et al.},
Phys.\ Lett.\ B {\bf 328}, 234 (1994);\\
P.~Anselmann {\it et al.}  [GALLEX Collaboration.],
Phys.\ Lett.\ B {\bf 342}, 440 (1995);\\
Y.~Fukuda {\it et al.}  [Super-Kamiokande Collaboration],
Phys.\ Rev.\ Lett.\ {\bf 82}, 1810 (1999).

%
\bibitem{seesaw} T.~Yanagida, in {\it Proceedings of the Workshop on Unified
Theory and Baryon Number of the Universe}, Tsukuba, Japan, 1979, edited
by O.~Sawada and A.~Sugamoto (KEK, Tsukuba, 1979); \\
M.~Gell-Mann, P.~Ramond, and R.~Slansky, in {\it Supergravity}, Proceedings
of the Workshop, Stony Brook, NY, 1979, edited by P.~van Nieuwenhuizen
and D.~Freedman (North-Holland, Amsterdam, 1979).
%
\bibitem{Borzumati:1986qx}
F.~Borzumati and A.~Masiero,
Phys.\ Rev.\ Lett.\ {\bf 57}, 961 (1986);\\
J.~Hisano, T.~Moroi, K.~Tobe, M.~Yamaguchi and T.~Yanagida,
Phys.\ Lett.\ B {\bf 357}, 579 (1995)
[hep-ph/9501407].

\bibitem{Hisano:1996cp}
J.~Hisano, T.~Moroi, K.~Tobe and M.~Yamaguchi,
Phys.\ Rev.\ D {\bf 53}, 2442 (1996)
[hep-ph/9510309].

\bibitem{Sato:2001zh}
J.~Sato, K.~Tobe and T.~Yanagida,
Phys.\ Lett.\ B {\bf 498}, 189 (2001)
[hep-ph/0010348];\\
J.~Sato and K.~Tobe,
hep-ph/0012333.

\bibitem{Hisano:1999fj}
J.~Hisano and D.~Nomura,
Phys.\ Rev.\ D {\bf 59}, 116005 (1999)
[hep-ph/9810479];hep-ph/0004061.

\bibitem{others}
J.~Hisano, D.~Nomura and T.~Yanagida,
Phys.\ Lett.\ B {\bf 437}, 351 (1998)
[hep-ph/9711348];\\
J.~Ellis, M.~E.~Gomez, G.~K.~Leontaris, S.~Lola and D.~V.~Nanopoulos,
Eur.\ Phys.\ J.\ C {\bf 14}, 319 (2000) [hep-ph/9911459];\\
J.~L.~Feng, Y.~Nir and Y.~Shadmi,
Phys.\ Rev.\ D {\bf 61}, 113005 (2000) [hep-ph/9911370];\\
W.~Buchm\"uller, D.~Delepine and L.~T.~Handoko,
Nucl.\ Phys.\ B {\bf 576}, 445 (2000) [hep-ph/9912317].

\bibitem{Brown:2001mg}
H.~N.~Brown {\it et al.}  [Muon g-2 Collaboration],
hep-ex/0102017.

\bibitem{Czarnecki:2001pv}
A.~Czarnecki and W.~J.~Marciano,
hep-ph/0102122.

\bibitem{g-2_before}
R.~Barbieri and L.~Maiani,
Phys.\ Lett.\ B {\bf 117}, 203 (1982);\\
D.~A.~Kosower, L.~M.~Krauss and N.~Sakai,
Phys.\ Lett.\ B {\bf 133}, 305 (1983);\\
T.~C.~Yuan, R.~Arnowitt, A.~H.~Chamseddine and P.~Nath,
Z.\ Phys.\ C {\bf 26}, 407 (1984);\\
C.~Arzt, M.~B.~Einhorn and J.~Wudka,
Phys.\ Rev.\ D {\bf 49}, 1370 (1994)
[hep-ph/9304206];\\
J.~L.~Lopez, D.~V.~Nanopoulos and X.~Wang,
Phys.\ Rev.\ D {\bf 49}, 366 (1994)
[hep-ph/9308336];\\
U.~Chattopadhyay and P.~Nath,
Phys.\ Rev.\ D {\bf 53}, 1648 (1996)
[hep-ph/9507386];\\
T.~Moroi,
Phys.\ Rev.\ D {\bf 53}, 6565 (1996)
[hep-ph/9512396];\\
M.~Carena, G.~F.~Giudice and C.~E.~Wagner,
Phys.\ Lett.\ B {\bf 390}, 234 (1997)
[hep-ph/9610233].

\bibitem{g-2_recent}
L.~Everett, G.~L.~Kane, S.~Rigolin and L.~Wang,
hep-ph/0102145;\\
J.~L.~Feng and K.~T.~Matchev,
hep-ph/0102146;\\
E.A.~Baltz and P.~Gondolo,
hep-ph/0102147;\\
U.~Chattopadhyay and P.~Nath,
hep-ph/0102157;\\
S.~Komine, T.~Moroi and M.~Yamaguchi,
hep-ph/0102204.

\bibitem{g-2lfv1}
E.~Ma and M.~Raidal,
hep-ph/0102255.

\bibitem{g-2lfv2}
T.~Huang, Z.~H.~Lin, L.~Y.~Shan and X.~Zhang,
hep-ph/0102193.


\bibitem{Ellis:1984bm}
J.~Ellis, C.~Kounnas and D.~V.~Nanopoulos,
Nucl.\ Phys.\ B {\bf 247}, 373 (1984);\\
A.~B.~Lahanas and D.~V.~Nanopoulos,
Phys.\ Rep.\ {\bf 145}, 1 (1987).

\bibitem{Kaplan:2000ac}
D.~E.~Kaplan, G.~D.~Kribs and M.~Schmaltz,
Phys.\ Rev.\ D {\bf 62}, 035010 (2000)
[hep-ph/9911293];\\
Z.~Chacko, M.~A.~Luty, A.~E.~Nelson and E.~Ponton,
JHEP {\bf 0001}, 003 (2000);\\
M.~Schmaltz and W.~Skiba,
Phys.\ Rev.\ D {\bf 62}, 095005 (2000)
[hep-ph/0001172].

\bibitem{Brooks:1999pu}
M.~L.~Brooks {\it et al.}  [MEGA Collaboration],
Phys.\ Rev.\ Lett.\ {\bf 83}, 1521 (1999)
[hep-ex/9905013].

\bibitem{Hisano:1997qq}
J.~Hisano, T.~Moroi, K.~Tobe and M.~Yamaguchi,
Phys.\ Lett.\ B {\bf 391}, 341 (1997)
[hep-ph/9605296].

\bibitem{Okada:2000zk}
Y.~Okada, K.~Okumura and Y.~Shimizu,
Phys.\ Rev.\ D {\bf 61}, 094001 (2000)
[hep-ph/9906446].

\bibitem{effsusy}
M.~Dine, A.~Kagan and S.~Samuel,
Phys.\ Lett.\ B {\bf 243}, 250 (1990);\\
S.~Dimopoulos and G.~F.~Giudice,
Phys.\ Lett.\ B {\bf 357}, 573 (1995)
[hep-ph/9507282];\\
A.~Pomarol and D.~Tommasini,
Nucl.\ Phys.\ B {\bf 466}, 3 (1996)
[hep-ph/9507462];\\
A.~G.~Cohen, D.~B.~Kaplan and A.~E.~Nelson,
Phys.\ Lett.\ B {\bf 388}, 588 (1996)
[hep-ph/9607394];\\
J.~Hisano, K.~Kurosawa and Y.~Nomura,
Phys.\ Lett.\ B {\bf 445}, 316 (1999)
[hep-ph/9810411];
Nucl.\ Phys.\ B {\bf 584}, 3 (2000)
[hep-ph/0002286].

\bibitem{Nilles:1984ge}
For a review, see H.~P.~Nilles,
%``Supersymmetry, Supergravity And Particle Physics,''
Phys.\ Rep.\ {\bf 110}, 1 (1984).

\bibitem{gmsb}
M.~Dine, A.~E.~Nelson, Y.~Nir and Y.~Shirman,
Phys.\ Rev.\ D {\bf 53}, 2658 (1996)
[hep-ph/9507378];\\
for a review,
see G.~F.~Giudice and R.~Rattazzi,
%``Theories with gauge-mediated supersymmetry breaking,''
Phys.\ Rep.\ {\bf 322}, 419 (1999)
[hep-ph/9801271].

\bibitem{neutrino_models1} 
J.~Sato and T.~Yanagida,
Phys.\ Lett.\ B {\bf 430}, 127 (1998);\\
C.~H.~Albright, K.~S.~Babu and S.~M.~Barr,
Phys.\ Rev.\ Lett.\ {\bf 81}, 1167 (1998)
[hep-ph/9802314];\\
N.~Irges, S.~Lavignac and P.~Ramond,
Phys.\ Rev.\ D {\bf 58}, 035003 (1998)
[hep-ph/9802334];\\
W.~Buchm\"uller and T.~Yanagida,
Phys.\ Lett.\ B {\bf 445}, 399 (1999)
[hep-ph/9810308].

\bibitem{JY} J.~Sato and T.~Yanagida, hep-ph/0009205.

\bibitem{neutrino_models2} For examples,
C.~H.~Albright and S.~M.~Barr,
Phys.\ Rev.\ D {\bf 62}, 093008 (2000)
[hep-ph/0003251];\\
K.~Hagiwara and N.~Okamura,
Nucl.\ Phys.\ B {\bf 548}, 60 (1999)
[hep-ph/9811495];\\
M.~Bando, T.~Kugo and K.~Yoshioka,
Phys.\ Lett.\ B {\bf 483}, 163 (2000)
[hep-ph/0003231];\\
G.~Altarelli, F.~Feruglio and I.~Masina,
hep-ph/0007254;\\
Y.~Nomura and T.~Yanagida,
Phys.\ Rev.\ D {\bf 59}, 017303 (1999)
[hep-ph/9807325].

\bibitem{Komine:2001tj}
S.~Komine and M.~Yamaguchi,
Phys.\ Rev.\ D {\bf 63}, 035005 (2001)
[hep-ph/0007327].

\bibitem{komine}
S.~Komine,
hep-ph/0102030;\\
S.~Komine, T.~Moroi and M.~Yamaguchi,
hep-ph/0103182.

\bibitem{PSI} L.M.~Barkov {\it et al.}, Research Proposal to PSI, 1999.
See also http://www.icepp.s.u-tokyo.ac.jp/meg/.
%
\bibitem{MECO} M.~Bachman {\it et al.} [MECO Collaboration],
Proposal to BNL, 1997. See also http://meco.ps.uci.edu.
%
\bibitem{CERN} See the WEB page of ``Neutrino factory and
muon storage rings at CERN'';
http://muonstoragerings.web.cern.ch/muonstoragerings/.
%
\bibitem{PRISM} See technical notes in the homepage of the
PRISM project: http://www-prism.kek.jp.
%
\end{thebibliography}
\end{document}